\documentclass[a4paper,fleqn]{cas-sc}

\usepackage[numbers]{natbib}

\usepackage{algorithm}
\usepackage{algpseudocode}
\usepackage{lineno}
\usepackage{subcaption}
\usepackage{siunitx}

\def\tsc#1{\csdef{#1}{\textsc{\lowercase{#1}}\xspace}}
\tsc{WGM}
\tsc{QE}

\begin{document}
\let\WriteBookmarks\relax
\def\floatpagepagefraction{1}
\def\textpagefraction{.001}

\shorttitle{}    

\shortauthors{X. Xie et~al.}  

\title [mode = title]{Improved Pixel-wise Calibration for Charge-Integrating Hybrid Pixel Detectors with Performance Validation}



%

\author[a]{X.~Xie} [
    orcid=0000-0001-6473-7886,]
\cormark[1]
\cortext[1]{Corresponding author}
\ead{xiangyu.xie@psi.ch}
\author[a]{A.~Bergamaschi} \author[a]{M.~Br\"uckner} \author[a]{M.~Carulla} \author[a]{R.~Dinapoli} \author[a]{S.~Ebner} \author[a]{K.~Ferjaoui}  \author[a]{E.~Fr\"ojdh} \author[a]{V.~Gautam} \author[a]{D.~Greiffenberg} \author[a]{S.~Hasanaj} \author[a]{J.~Heymes} \author[a]{V.~Hinger} \author[a]{M.~H\"urst} \author[a]{V.~Kedych} \author[a]{T.~King} \author[a]{S.~Li} \author[a]{C.~Lopez-Cuenca} \author[a]{A.~Mazzoleni}  \author[a]{D.~Mezza} \author[a]{K.~Moustakas} \author[a]{A.~Mozzanica} \author[a]{J.~Mulvey} \author[a]{M.~M\"uller} \author[a]{K.A.~Paton} \author[a]{C.~Posada~Soto} \author[a]{C.~Ruder} \author[a]{B.~Schmitt} \author[a]{P.~Sieberer} \author[a]{S.~Silletta} \author[a]{D.~Thattil} \author[a]{~J.~Zhang}




\affiliation[a]{organization={PSI Center for Photon Science},
            postcode={5232 Villigen PSI},  
            country={Switzerland}}




\begin{abstract}
  The MÖNCH hybrid pixel detector, with a 25~µm pixel pitch and fast charge-integrating readout, has demonstrated subpixel resolution capabilities for X-ray imaging and deep learning--based electron localization in electron microscopy.
  Fully exploiting this potential requires extensive calibration to ensure both linearity and uniformity of the pixel response, which is challenging for detectors with a large dynamic range.
  To overcome the limitations of conventional calibration methods, we developed an accurate and efficient correction method to achieve pixel-wise gain and nonlinearity calibration based on the backside pulsing technique.
  A three-dimensional lookup table was generated for all pixels across the full dynamic range, mapping the pixel response to a calibrated linear energy scale.

  Compared with conventional linear calibration, the proposed method yields negligible deviations between the calibrated and nominal energies for photons and electrons.
  The improvement in energy resolution ranges from 4\% to 22\% for 15--25 keV photons and from 16\% to 22\% for 60--200 keV electrons.
  Deep learning--based electron localization demonstrates a 4\% improvement in spatial resolution when using the proposed calibration method.
  This approach further enables rapid diagnosis of the cause of bad pixels and estimation of bump-bonding yield.

\end{abstract}





\begin{keywords}
 Hybrid pixel detector \sep
 Charge-integrating readout \sep
 Detector calibration \sep
 X-ray detectors
\end{keywords}

\maketitle


\section{Introduction}\label{}
The MÖNCH detector \cite{MONCH}, a charge-integrating hybrid pixel detector with a 25~µm pixel pitch and fast frame rate up to 6 kHz, has demonstrated capabilities for data enhancement such as subpixel resolution in X-ray imaging \cite{Chiriotti_2022} and deep learning--based electron localization with improved position resolution in electron microscopy \cite{EM_deepLearning2023}.
These applications will further benefit from improved detector performance, particularly better energy resolution and pixel-to-pixel uniformity, both of which strongly affect the achievable spatial resolution.

Previous calibration studies for charge-integrating detectors such as AGIPD \cite{Mezza_2016} and JUNGFRAU \cite{Mozzanica_2016} have adopted constant gains and offsets assuming a perfectly linear response and focused on determining switching points between different gain modes in their adaptive gain design.
In low-occupancy ($\sim 1\%$) scenarios, however, a single gain stage is typically used for readout.
In such cases, the calibration should focus on the linearity of the pixel response and on uniformity across pixels to improve the energy resolution.
In practice, the pixel response exhibits intrinsic nonlinearity arising from the readout chain, including the amplifier, readout buffer, and analog-to-digital converter (ADC), while non-uniformity primarily originates from variations in the amplifiers, column buffers, and ADC channels \cite{Dinapoli_2014}.

The conventional calibration method for charge-integrating detectors commonly uses X-ray fluorescence sources.
However, the limited number of available X-ray fluorescence energies, along with the overlap of fluorescence lines, makes it difficult to characterize the nonlinearity of the pixel response.
Furthermore, the low flux of these sources, together with the limited stopping power of sensors for higher-energy photons, results in long calibration times to accumulate sufficient statistics for pixel-wise calibration and in limited capability to cover the dynamic range of the detector.
Furthermore, the small \si{25~\micro\meter} pixel pitch of MÖNCH makes it highly likely that the charge generated by a single photon will be shared among neighboring pixels \cite{XIE2026170894}.
The resulting fractional energy collected by a single pixel, combined with the pixel response nonlinearity, further complicates calibration.

Synchrotron radiation sources provide higher flux and monochromatic photons, but their accessibility is limited.
As summarized in \cite{MEZZA2022166078}, the on-chip current source offers a convenient way to calibrate all pixels simultaneously, but its accuracy is limited by non-uniformities arising from power supply variations, instability, manufacturing non-uniformity, and pedestal drift during exposure time scans.
Although pulsed lasers overcome these limitations, their beam typically covers only a small detector area, making full calibration over the entire detector area almost impossible.
In addition, the method requires removal of the aluminium entrance window from the sensor for use with visible or infrared lasers, which is technically inconvenient.

In this work, we present an accurate and efficient calibration method for pixel-wise gain and nonlinearity correction based on the backside pulsing technique \cite{Mezza_2016}.
A three-dimensional lookup table (3D LUT) is generated for all pixels over the full dynamic range, mapping the pixel response in analog-to-digital unit (ADU) value to calibrated energy.
Measurements of a few million monochromatic X-ray photons across all pixels are used to determine the global energy conversion coefficient.
Compared with conventional linear calibration methods, the proposed approach significantly improves both energy accuracy and resolution for X-ray photons and electrons.
The calibration takes less than two hours independent of pixel array size, along with only a limited dependence on the X-ray source.

This paper is organized as follows.
Section~\ref{sec:procedure} describes the calibration and analysis procedures in detail.
Results are presented in Section~\ref{sec:results}, including examinations of the nonlinearity and non-uniformity of the pixel response, energy spectra for photons and electrons at different energies, and comparative deep-learning results obtained with different calibration methods.
Section~\ref{sec:discussion} discusses the calibration method, its potential applications in pixel yield diagnosis, and its practical implementation.
Finally, Section~\ref{sec:conclusion} summarizes the work.

\section{Calibration procedure}
\label{sec:procedure}

\subsection{MÖNCH03 detector}
\begin{figure*}
    \centering
    \includegraphics[width=0.5\linewidth]{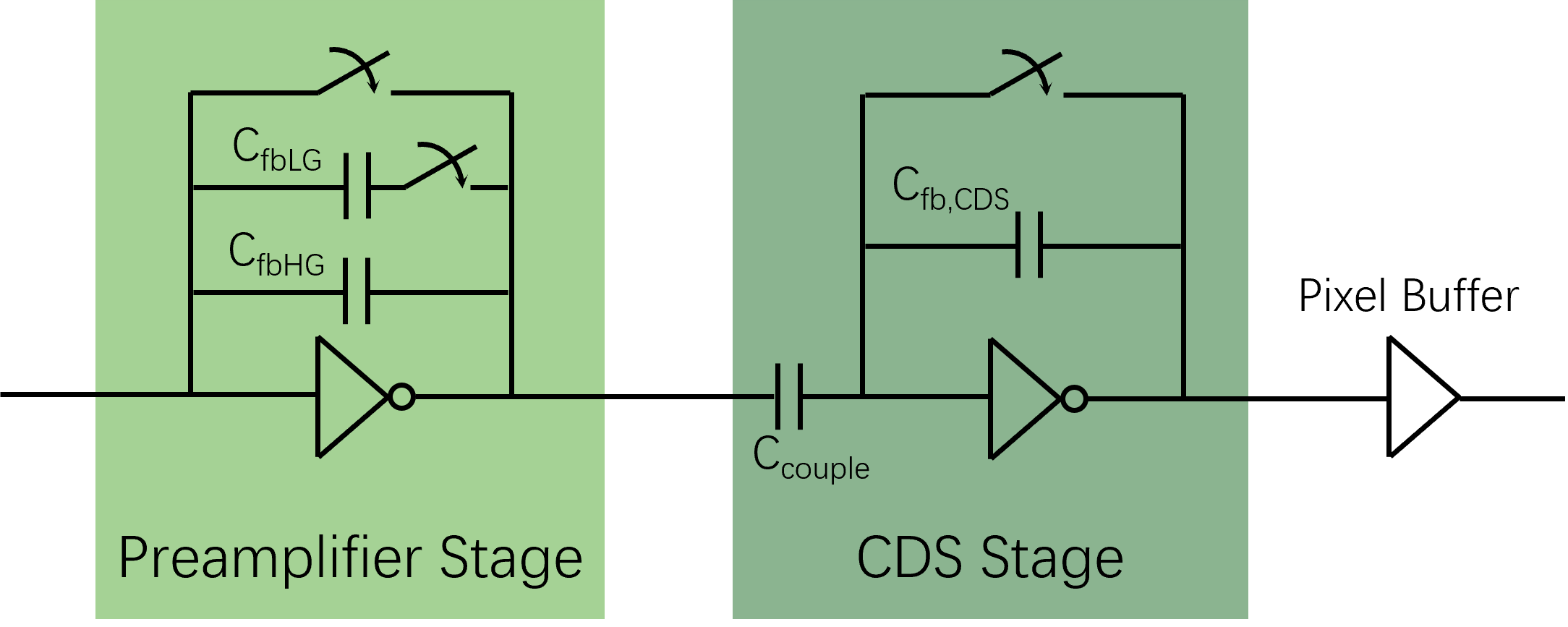}
    \caption{Simplified schematic of the pixel electronics in the MÖNCH ASIC, containing a charge-sensitive amplifier, a correlated double sampler (CDS), and configurable gain stages.}
    \label{fig:pixel_electronics}
\end{figure*}

The MÖNCH03 ASIC \cite{MONCH} under test has a $400 \times 400$ pixel array with a pixel pitch of 25~µm, bump-bonded to a 300~µm thick silicon sensor.
The pixel electronics, as detailed in Fig.~\ref{fig:pixel_electronics}, consist of a charge-sensitive amplifier to integrate charges within an exposure window, a correlated double sampler (CDS), and configurable gain stages: two selectable preamplifier gains (low gain and high gain) and three selectable CDS gains (g1, g2, and g4).
The analog outputs of every $25\times200$ pixel block are digitized by one 14-bit ADC channel.

\subsection{Calibration setup}
\label{sec:setup}

\begin{figure*}
    \centering
    \begin{subfigure}{0.6\linewidth}
      \centering
      \includegraphics[width=\linewidth]{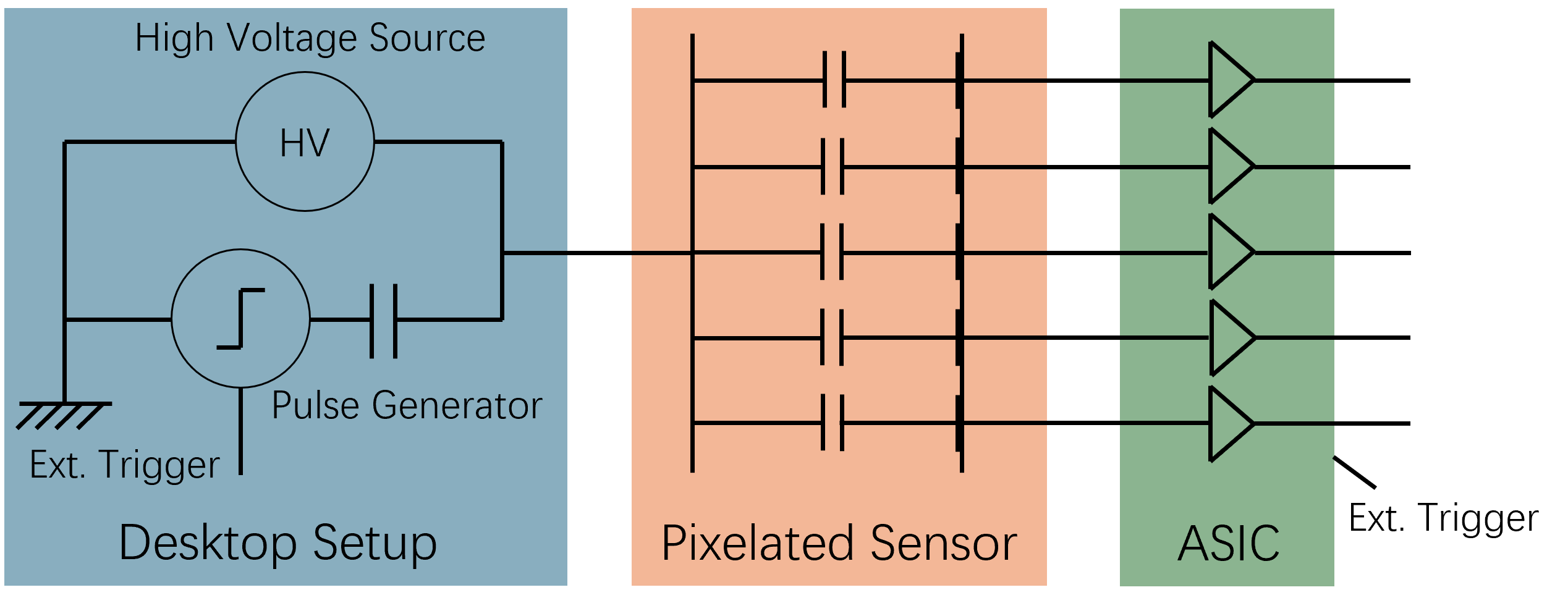}
      \subcaption{Calibration setup}
      \label{fig:setup}
    \end{subfigure}
    \begin{subfigure}{0.3\linewidth}
      \centering
      \includegraphics[width=\linewidth]{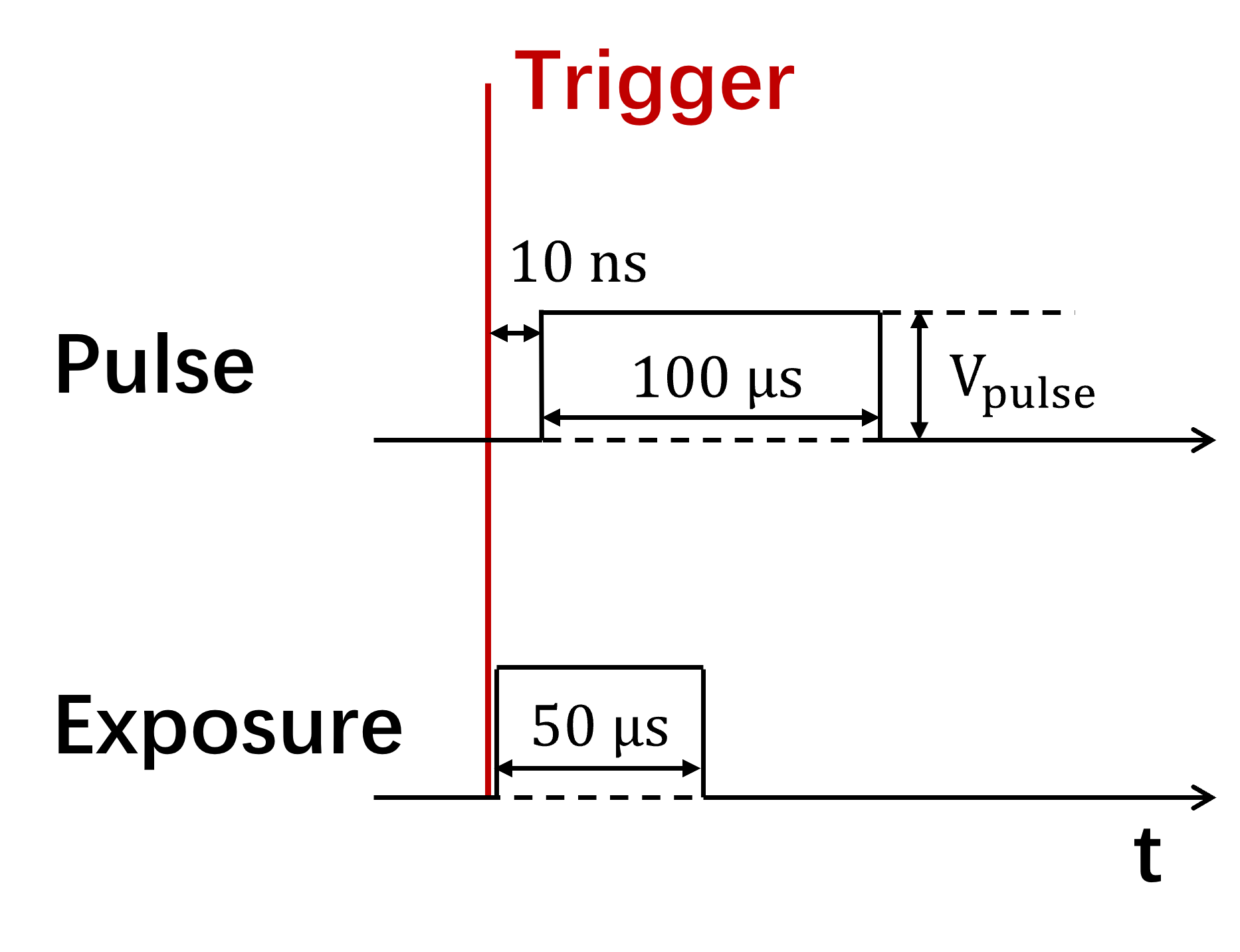}
      \subcaption{Timing structure}
      \label{fig:timing_structure}
      
    \end{subfigure}
    
    \caption{
      (a) Schematic of the calibration setup. 
      The high voltage from a Keithley 2410 is superimposed on the pulse generated by an Agilent 33250A to bias the backside of the MÖNCH sensor.
      The combined voltage is applied to the backside of the silicon sensor in the MÖNCH detector, where the signal is amplified by the pixel electronics in the ASIC.
      (b) Timing structure of the pulse and the MÖNCH detector exposure (not to scale).
      The pulse is synchronized with the MÖNCH detector exposure with a delay of 10~ns using an external 1~kHz trigger signal.
      The pulse duration is set to 100~\si{\micro\second}, larger than the 50~\si{\micro\second} exposure time but shorter than the 600~\si{\micro\second} readout period.
      This ensures that only the rising edge of the pulse is captured by the MÖNCH detector.
    }
\end{figure*}

Fig.~\ref{fig:setup} shows a schematic of the calibration setup.
A pulse generator (Agilent 33250A \cite{33250A}) produces a pulse with a precise amplitude $V_{\rm pulse}$\footnote{The accuracy is $\pm1\%$ of the setting, with an additional $\rm \pm 1\ mV_{pp}$.} via the high impedance output, which is coupled to the the backside of the sensor through a capacitor.
The sensor is biased to 90 V from a Keithley 2410 \cite{Keithley2410} power supply directly, sufficiently above the sensor depletion voltage of around 30~V.

In the data acquisition timing diagram shown in Fig.~\ref{fig:timing_structure}, an external 1 kHz TTL trigger signal synchronizes the pulse generator with the 50~µs exposure time of the MÖNCH detector.
The pulse generator is configured with a delay of 10~ns and a duration of 100~µs, ensuring that only the rising edge of the pulse is captured by the MÖNCH detector.
In this way, all pixels simultaneously receive an additional amount of charge 
\begin{equation}
  \label{eq:Q=CV}
  Q = C_{\rm pixel} \cdot V_{\rm pulse},
\end{equation}
where $C_{\rm pixel}$ is the pixel-to-backplane capacitance.
Under the planar-capacitance assumption and using Eq. \ref{eq:Q=CV}, the equivalent energy response of a pixel can be expressed as\footnote{As all pixels are pulsed at the same time, and the preamp DC gain is very high (>100), the interpixel capacitances can, to the first order, be neglected.}
\begin{equation}
  \label{eq:EnergyChargeConversion}
  E = \frac{\epsilon S}{d e}\cdot E_{i} \cdot V_{\rm pulse},
\end{equation}
where $\epsilon$ is the silicon permittivity, $S=25 \times 25~\si{\micro\meter\squared}$ is the pixel area, $d$ is the sensor thickness of 300~\si{\micro\meter}. 
$e$ is the elementary charge, and $E_{i}$ is the average electron--hole pair creation energy in silicon (3.62~eV -- 3.64~eV \cite{Eehpair_3_62eV, Eehpair_3_64eV}).
By scanning the pulse amplitude $V_{\rm pulse}$, the pixel response in ADU can be mapped as a function of the injected charge and the equivalent energy using Eqs.~\ref{eq:Q=CV} and \ref{eq:EnergyChargeConversion}.
The scan consisted of 300 pulse amplitudes on a linear scale from 0~V to 12~V (amplified by an integrated backside pulsing amplifier \cite{MEZZA2022166078}, not plotted in Fig.~\ref{fig:setup}) over the full dynamic range to ensure fine granularity.
For each pulse amplitude, 1000 frames were acquired both with and without the pulse to achieve sufficient statistics, taking approximately six seconds in total, including the waiting time.
The difference between the average readout in these two cases was used to determine the pixel response at the given pulse amplitude.
The scan was automated using a Python script and completed in one hour including setup time.

According to Eq.~\ref{eq:EnergyChargeConversion}, the energy conversion coefficient is approximately 4.9~keV/V for the MÖNCH03 detector under test.
However, the associated uncertainty is non-negligible\footnote{$E_{i}$ varies by about 1\% in the literature, the silicon permittivity depends on the signal frequency; and the sensor thickness may differ from the nominal value.}.
In practice, we determined this coefficient by measuring X-ray photons of known energy and aligning the measured energy centroid to the nominal photon energy.
Given the relatively small sensor thickness variation over an area of 1~\si{\centi\meter\squared}, we neglected variations in $C_{\rm pixel}$ and assumed that all pixels share the same coefficient.
For this reason, this measurement requires only about one million X-ray photons collected over the entire detector area.
It is also feasible to use laboratory X-ray fluorescence sources to determine the energy conversion coefficient, as demonstrated in Section~\ref{sec:analysis}.
Targets made of elements with well-separated K$\beta$ lines, such as tin (K$\beta_1$ at 28.486~keV, more than 3~keV apart from K$\alpha$ lines), are preferred.
Although the K$\alpha$ lines provide higher statistics than the K$\beta_1$ line, it is difficult to resolve the K$\alpha_1$ and K$\alpha_2$ components using MÖNCH03, and the energy centroid is unclear due to the unresolved K$\alpha_1$ and K$\alpha_2$ components, whose relative intensities can be affected by target geometry, thickness, and self-absorption effects.
Such a measurement can be acquired in a few seconds at a synchrotron radiation source or in a few minutes using laboratory X-ray fluorescence sources.

\subsection{Calibration data analysis}
\label{sec:analysis}

\begin{figure*}
  \centering
  \begin{subfigure}{0.45\linewidth}
    \centering
    \includegraphics[width=\linewidth]{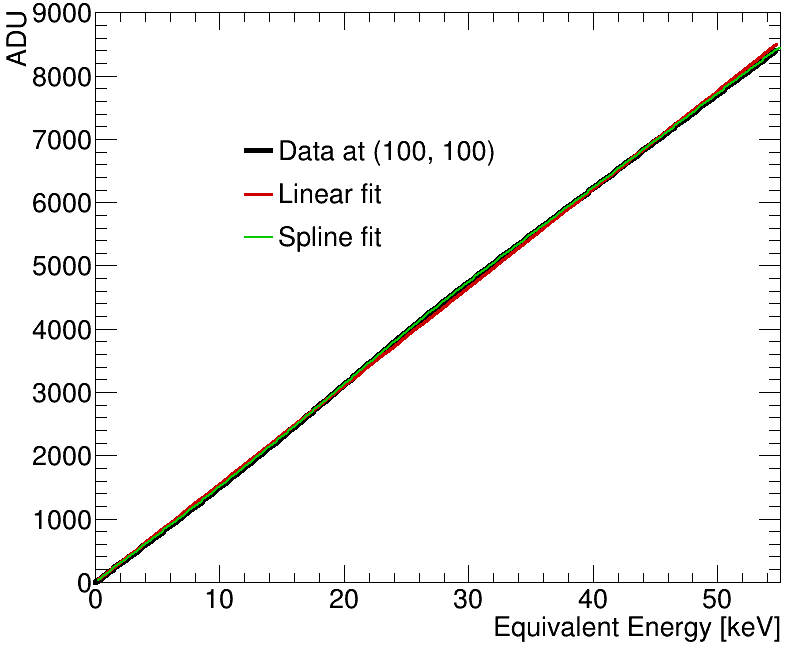}
    \subcaption{Pixel response and fits}
    \label{fig:response_and_fits}
  \end{subfigure}
  \ \ 
  \begin{subfigure}{0.45\linewidth}
    \centering
    \includegraphics[width=\linewidth]{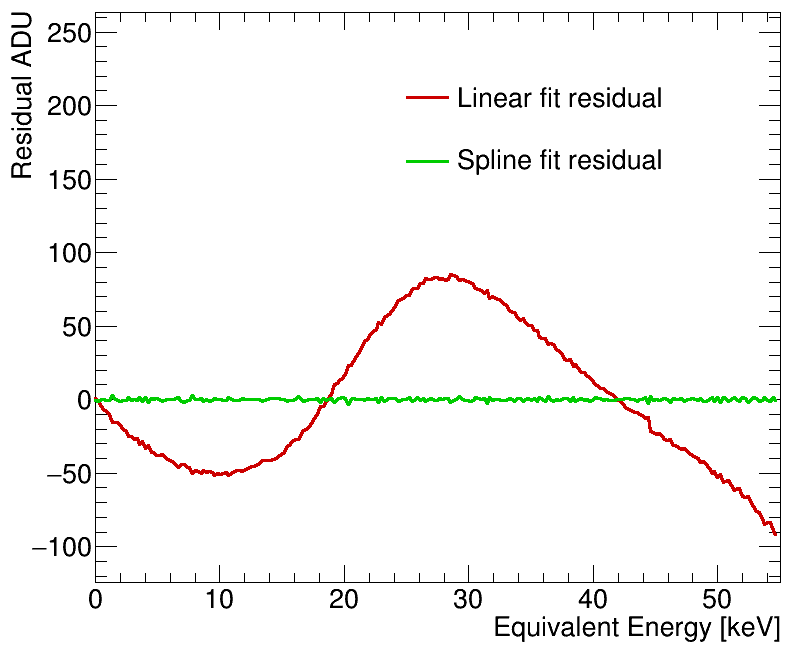}
    \subcaption{Residuals of the fits}
    \label{fig:residuals}
  \end{subfigure}
  \caption{
    (a) Pixel response at coordinates (100, 100) as a function of equivalent energy, along with linear and spline fits. The offset of the linear fit is constrained to zero.
    (b) Residuals of the measured pixel response with respect to the linear and spline fits.
  }
  \label{fig:response}
\end{figure*}

The response of a representative pixel at coordinates (100, 100) in the highest available gain setting is shown in Fig.~\ref{fig:response_and_fits} as a function of the equivalent energy, converted from the pulse amplitude as discussed above.
A linear fit (red) and a third-order spline fit (green) using the splrep function\footnote{The smoothing factor is $m + \sqrt{2m}$, where $m$ is the number of data points} from the SciPy package \cite{2020SciPy-NMeth} were applied to the pixel response.
The linear fit, with a gain of approximately 150~ADU/keV, deviates from the measured pixel response, varying along with the equivalent collected energy.
These differences are more evident in the residuals shown in Fig.~\ref{fig:residuals}.
In addition to small fluctuations due to noise, the residuals from the linear fit display a sine-like pattern with significant deviations up to about 0.7~keV.
The residual shape clearly reveals the nonlinearity of the pixel response, consistent with observations using discrete pile-up of 20~keV photons in \cite{Redford_2018}.

The spline fit, as expected, closely follows the measured response.
The chi-squared per degree of freedom ($\chi^2$/NDF) values for the linear and spline fits are 1876.8 and 0.8, respectively, indicating that the linear fit is not suitable for describing the pixel response while the spline fit provides a good approximation.
The RMS values of the residuals for the linear and spline fits are 62.6~ADU and 0.9~ADU, respectively, corresponding approximately to 0.42~keV and 0.01~keV.
Such large deviations and the resulting calibration biases would significantly impact the achievable energy accuracy, i.e., the closeness of the calibrated energy to the nominal energy, degrading MÖNCH's performance in energy-resolved applications.
This will be further discussed in Section~\ref{sec:discussion}.

Given the nonlinear pixel response, using a constant gain is not precise enough for calibration.
It is also impractical to construct a parametric model, since the exact functional form of the pixel response is undefined and varies across pixels.
To overcome these limitations, we adopt a three-dimensional lookup table (3D LUT) to map the ADU output of each pixel directly to the equivalent energy.
This lookup-based approach enables fast energy calibration with minimal computational overhead.
The spline fit shown in Fig.~\ref{fig:response_and_fits} was first performed for each pixel.
By inverting the spline fit and applying the subsequently determined energy conversion coefficient, we obtained a mapping from the pixel response to equivalent energy.

\begin{figure}
    \centering
    \includegraphics[width=0.45\linewidth]{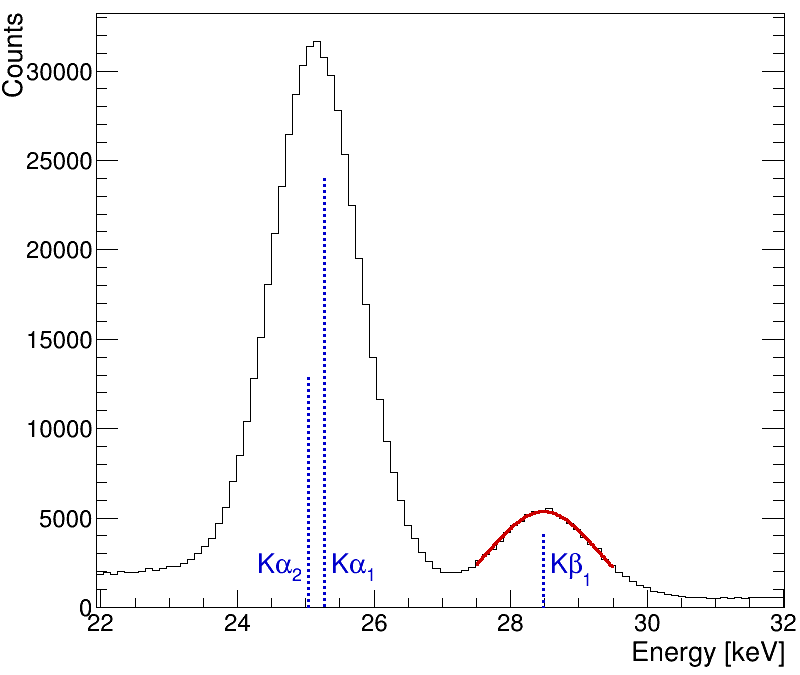}
    \caption{
      MÖNCH03 $3\times3$ cluster energy spectrum of Sn K$\alpha$ and K$\beta$ fluorescence photons after the 3D LUT calibration, where the energy conversion coefficient was determined using synchrotron monochromatic X-rays. 
      The Sn K$\alpha$ and K$\beta$ lines where the line energies and theoretical relative intensities are taken from \cite{XrayBooklet, SCOFIELD1974121}.
      A Gaussian fit (shown in red) to the Sn K$\beta$ peak yields an energy centroid of 28.484~keV, well aligned with the nominal energy of 28.486~keV \cite{XrayBooklet}.
      The good agreement between the Sn K$\beta$ line with the nominal energy indicates the accuracy of the proposed calibration method and demonstrates that laboratory X-ray fluorescence sources can be used to determine the energy conversion coefficient.
    }
    \label{fig:SnKbeta}
\end{figure}

To obtain the energy conversion coefficient for this detector, we used monochromatic X-ray photons collected at the METROLOGIE beamline \cite{METROLOGIE} of the SOLEIL synchrotron.
Alternatively, laboratory X-ray fluorescence sources can be used, as discussed in Section~\ref{sec:setup}.
Fig.~\ref{fig:SnKbeta} shows the Sn fluorescence spectrum measured by the MÖNCH03 detector.
A Gaussian fit to the Sn K$\beta_1$ peak\footnote{For Sn K$\alpha$, it is hard to determine its centroid due to the mixed Sn K$\alpha_1$ (25.271~keV) and K$\alpha_2$ (25.044~keV) lines \cite{XrayBooklet} with unknown relative intensities, whereas K$\beta_1$ (28.486~keV) is well separated using MÖNCH03.} yields an energy centroid of 28.484~keV with good fitting quality, well aligned with the nominal energy of 28.486~keV \cite{XrayBooklet}.
The high consistency indicates the accuracy of the proposed calibration method.
This result also demonstrates the capability of using laboratory X-ray fluorescence sources to determine the energy conversion coefficient.

We applied special treatment for pixels at the detector edges, which have larger capacitance due to the narrow guard ring design and therefore a larger response as described by Eq.~\ref{eq:Q=CV}.
The calibration data for the outermost 15 pixels were replaced with the average of their inner neighboring pixels read out by the same ADC.
In the future, we are going to optimize the guard ring structure to make the pixel-to-backplane capacitance consistent over the entire active area.

The output of the detector is sampled by 14-bit ADCs (0 to 16,383~ADU).
The full range was rebinned into 1,639 bins (10 ADU/bin) to balance granularity and LUT size.
Interpolation between LUT entries can be performed to preserve the energy resolution.
An additional LUT bin was reserved to flag pixels with failed spline fits, typically due to dead or noisy readout channels.
For the 400~$\times$~400-pixel MÖNCH03 detector, the resulting LUT size is 1.0~GB in single-precision floating-point format.

\section{Results}
\label{sec:results}

\subsection{Pixel response nonlinearity and non-uniformity}
\label{sec:nonlinearity}
\begin{figure*}
    \centering
    \begin{subfigure}{0.45\linewidth}
      \centering
      \includegraphics[width=\linewidth]{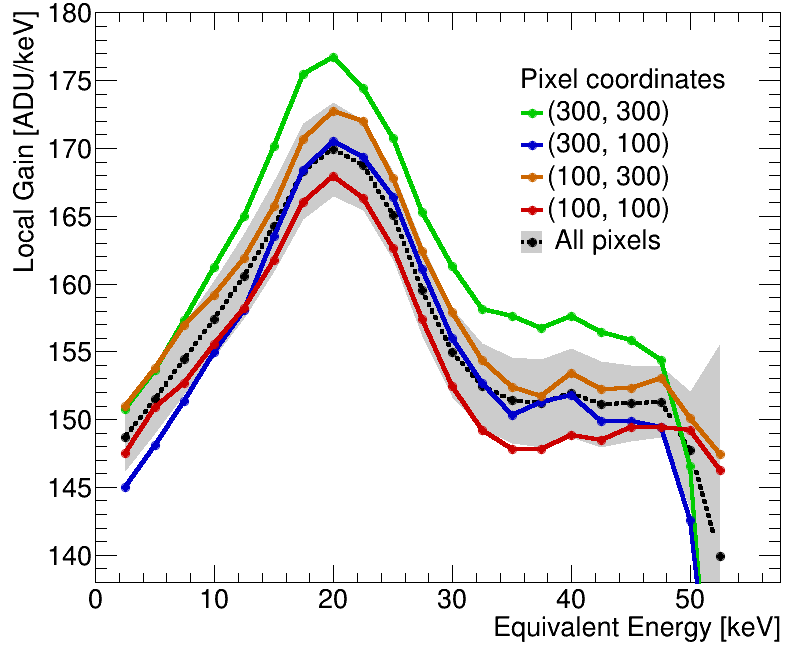}
      \subcaption{Local gain as a function of equivalent energy}
      \label{fig:local_gain}
    \end{subfigure}
    \ \ 
    \begin{subfigure}{0.45\linewidth}
      \centering
      \includegraphics[width=\linewidth]{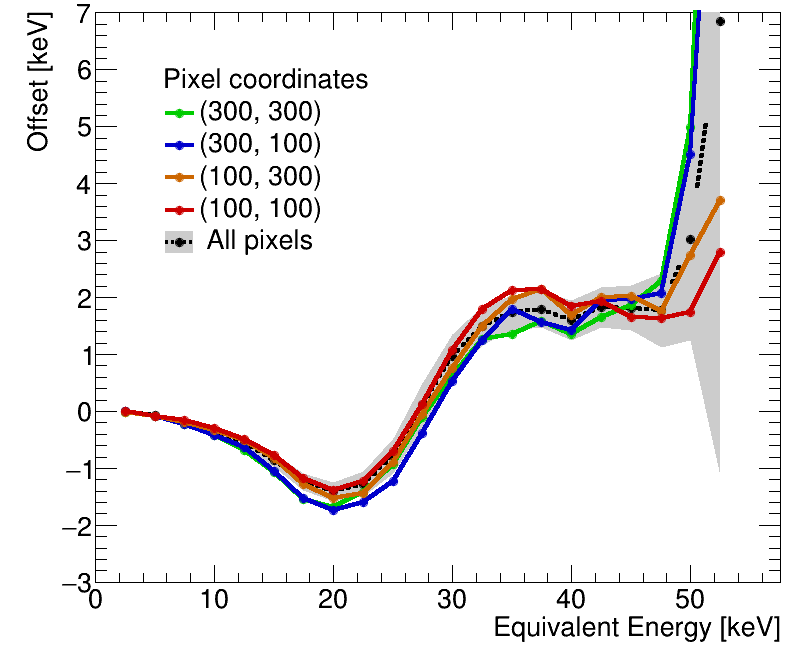}
      \subcaption{Offset as a function of equivalent energy} 
      \label{fig:local_offset}
    \end{subfigure}
    \caption{
      (a) Local gain, i.e., the derivative of the ADU--keV response within a 5~keV window; (b) offset as a function of equivalent energy for different pixels, and averaged over all pixels with error bars representing the standard deviation across pixels.
      The results are obtained by performing linear fits over a 5~keV sliding window, sampled every 2.5~keV
    }
\end{figure*}
Based on the 3D LUT calibration data, we investigated the nonlinearity and non-uniformity of the pixel response as would be obtained using the conventional linear calibration method.
At each 2.5~keV step across the energy range, local gains and offsets were extracted from linear fits using two ADU--energy points separated by 5~keV across the dynamic range, mimicking the limited energy ranges typically available from fluorescence sources.
The local gains of four representative pixels as functions of the equivalent energy are shown in Fig.~\ref{fig:local_gain}.
For individual pixels, this local gain varies significantly with energy, with maximum deviations exceeding 20\%.  
In addition, the average local gain over all pixels is shown, with error bars representing the standard deviation across pixels, revealing substantial pixel-to-pixel variations.

Fig.~\ref{fig:local_offset} presents the corresponding offsets, converted to keV using the fitted gain.  
At higher energies, the offset is typically non-zero and often exceeds 1~keV.  
Introducing a non-zero offset reduces the residuals between the fit and the measured response, but it also implies a non-zero calibrated output for zero input, which has no physical meaning.

These results demonstrate that pixel responses exhibit strong nonlinearity.
Extrapolating a local linear fit beyond the fitted window would lead to large calibration errors.  
Although the overall trends of the gain and offset curves are similar across pixels, noticeable pixel-to-pixel variations in both amplitude and shape remain, reflecting the non-uniformity across the pixel array.
Both the nonlinearity and non-uniformity effects degrade the achievable energy resolution when conventional linear calibration is applied.


\subsection{Calibrated photon energy spectra}
\label{sec:photon_results}
The performance of different calibration methods was first evaluated using photon energy spectra.
We conducted measurements with monochromatic X-ray photons at the METROLOGIE beamline \cite{METROLOGIE} of the SOLEIL synchrotron.
We used a low flux X-ray beam incident perpendicularly to the sensor surface of a MÖNCH03 detector ($\sim 1\%$ occupancy) operated in the highest gain.
The detector exposure time was set to the same 50~µs as used in the calibration procedure with the silicon sensor biased at 90~V.
The beam energy was set to 15~keV, 20~keV, and 25~keV, respectively, with the beam covering more than 10,000 pixels.

The 3D LUT calibration method was applied as described above for the MÖNCH detector.
Conventional methods were emulated by performing zero-offset linear fits to the full range of the 3D LUT at two levels of granularity: a pixel-wise fit and a global fit.
Although it is hard to achieve pixel-wise linear calibration in practice due to the charge sharing effect, we included this method in the comparison to demonstrate the impact of non-uniformity across pixels.

The raw data were processed using different calibration methods.
Given the 25~µm pixel pitch, we used $3\times3$ pixel clusters with the central pixel having the largest signal to capture the full energy deposition of a single photon \cite{Cartier_2014}.
Events affected by dead or noisy pixels were excluded from the analysis using the last LUT bin reserved for flagging such pixels.
Pile-up events were removed by requiring that no pixel in the concentric regions of a hit, between the $3\times3$ and $7\times7$ pixel areas, had a readout value exceeding three times its noise level.

\begin{figure*}
    \centering
    \includegraphics[width=0.6\textwidth]{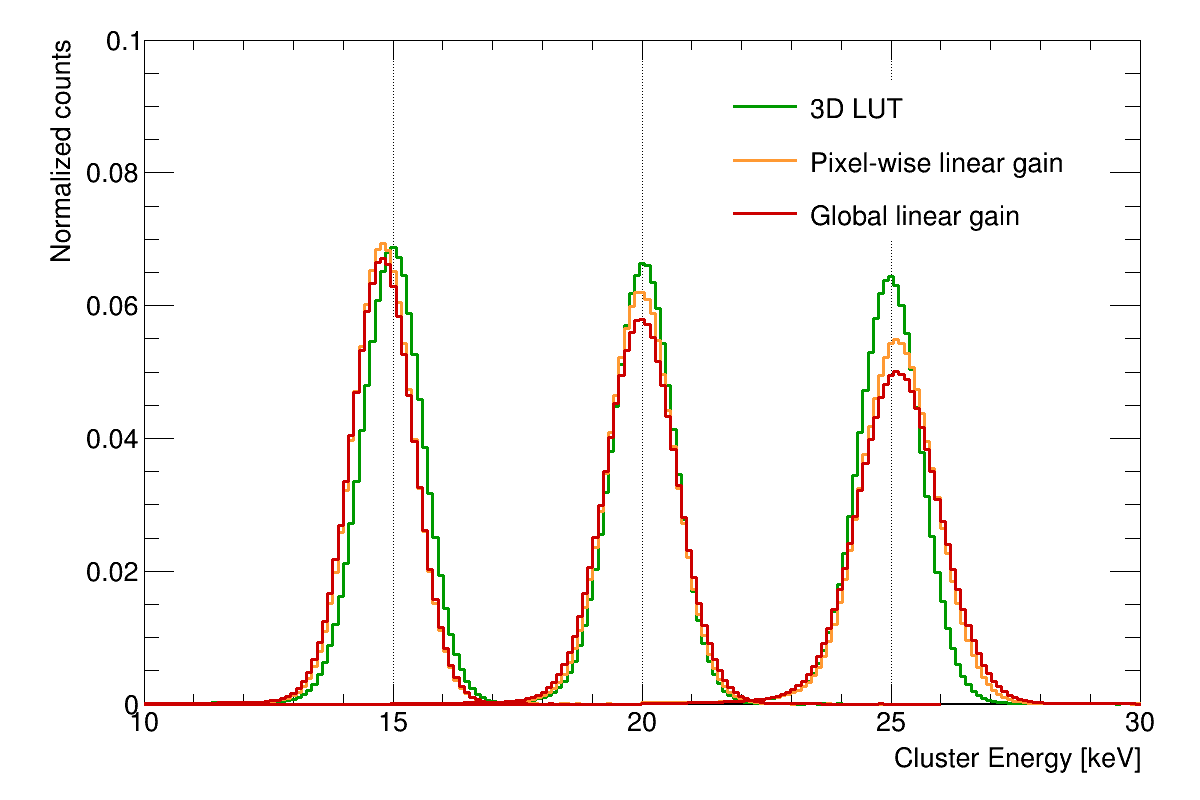}
    \caption{
      Monochromatic photon energy spectra for 15~keV, 20~keV, and 25~keV X-rays, acquired with different calibration methods, using the measurements carried out at the METROLOGIE beamline of the SOLEIL synchrotron.
    }
    \label{fig:spectrum}
\end{figure*}

The normalized energy spectra of 15~keV, 20~keV, and 25~keV X-rays obtained with different calibration methods are shown in Fig.~\ref{fig:spectrum}.
The 3D LUT calibration results exhibit both good alignment with the nominal photon energies and narrow energy spread.
In contrast, spectra obtained with pixel-wise linear calibration show noticeable deviations and poorer resolutions, while the resolutions of the global linear calibration degrade further, as expected.

Gaussian fits were applied to the spectra to extract the energy centroid and the standard deviation.
The energy centroid reflects the calibration accuracy, while the ratio of the full width at half maximum (FWHM=$2.355\times\sigma_E$) to the centroid (FWHM/$E_{\rm centroid}$) quantifies the relative energy resolution.
These results are quantified and summarized in Table \ref{tab:photon_results}.
As the photon energy increases, the relative energy resolution generally improves due to the higher signal-to-noise ratio (SNR).
For 15~keV X-rays, the relative energy resolution obtained with the pixel-wise linear calibration is comparable to that of the 3D LUT method.
This is because the pixel response is approximately linear below 10~keV (as shown in Fig.~\ref{fig:response_and_fits}), where most of the pixel energies lie for 15~keV photons.
For the 20~keV and 25~keV X-rays, however, the 3D LUT calibration achieves relative energy resolutions of 7.28\% and 5.89\%, respectively, significantly better than the 8.33\% and 7.56\% obtained with pixel-wise linear calibration.
These improvements arise from the more pronounced nonlinearity of the pixel response in the wider energy range.
In general, the 3D LUT calibration method outperforms the conventional linear calibration methods with improvements of 4\%, 13\%, and 22\% in relative energy resolution for 15~keV, 20~keV, and 25~keV photons, respectively.

\begin{table*}
  \caption{Summary of the photon energy spectra in terms of the energy centroid ($E_{\rm centroid}$) and relative energy resolution (FWHM/$E_{\rm centroid}$) for different photon energies using different calibration methods.}
  \begin{tabular}{c|c|c|c|c|c|c}
    \bf{Fitted results} & \multicolumn{3}{c|}{\bf{$\boldsymbol{E_{\text{centroid}}}$ [keV]}} & \multicolumn{3}{c}{\bf{FWHM/$\boldsymbol{E_{\text{centroid}}}$ [\%]}} \\
    \hline
    \bf{Photon energy [keV]} & \bf{15} & \bf{20} & \bf{25} & \bf{15} & \bf{20} & \bf{25} \\ 
    \hline
    \bf{3D LUT calibration} & 14.99 & 20.03 & 24.98 & 9.54 & 7.28 & 5.89 \\
    \hline
    \bf{Pixel-wise linear gain} & 14.81 & 20.01 & 25.16 & 9.56 & 7.77 & 6.93 \\
    \hline
    \bf{Global linear gain} & 14.79 & 19.99 & 25.14 & 9.91 & 8.33 & 7.56 \\
  \end{tabular}
  \label{tab:photon_results}
\end{table*}

\subsection{Calibrated electron energy spectra}
\label{sec:electron_results}
Electron measurements pose additional challenges compared to photon measurements, as electrons travel through the sensor due to multiple scattering, depositing energy over a larger area and across a wider energy range.
Electrons with energies of 60~keV, 80~keV, and 200~keV were collected using a MÖNCH03 detector installed on a JEOL JEM-ARM200F NEOARM electron microscope at the Electron Microscopy Facility of the Paul Scherrer Institute (PSI).  
No sample was inserted in the microscope.
The MÖNCH03 detector was bottom-mounted using custom-built mechanics and operated in the middle gain mode.
Calibration for this detector was conducted as described in Section~\ref{sec:procedure}.

\begin{figure*}
    \centering
    \includegraphics[width=0.8\textwidth]{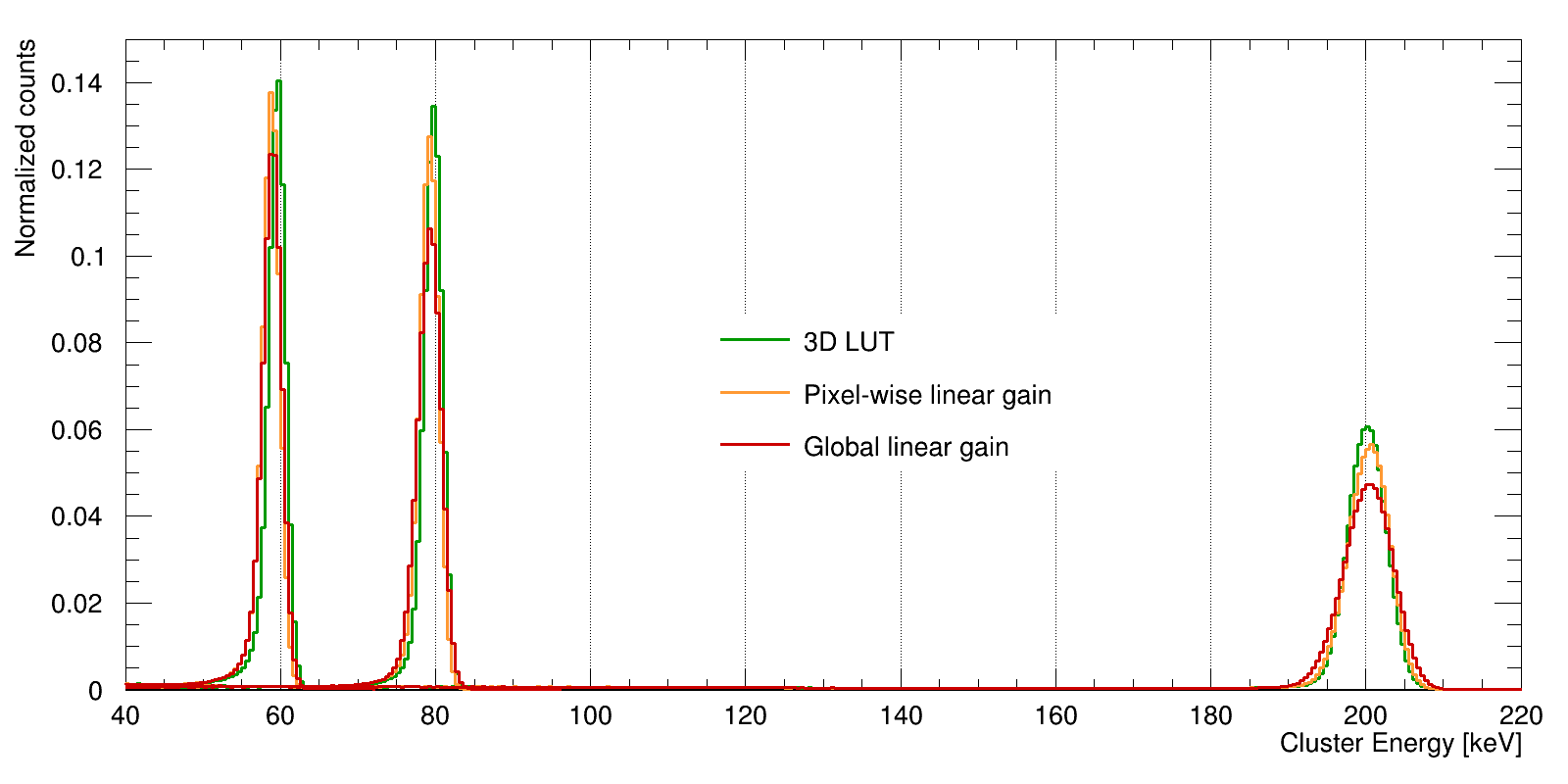}
    \caption{
      Electron energy spectra for 60~keV, 80~keV, and 200~keV electrons obtained with different calibration methods, using a JEOL JEM-ARM200F NEOARM electron microscope at the Electron Microscopy Facility, Paul Scherrer Institute (PSI).
    }
    \label{fig:electron_spectrum}
\end{figure*}

Electron clusters were formed by first selecting a seed pixel with a signal exceeding five times its noise level and then recursively adding neighboring pixels with signals exceeding two times the noise level.
The normalized electron energy spectra at 60~keV, 80~keV, and 200~keV obtained using different calibration methods are shown in Fig.~\ref{fig:electron_spectrum}.
Similar to the photon results, the 3D LUT calibration method achieves negligible bias and the narrowest energy spread.
As summarized in Table~\ref{tab:electron_results}, the 3D LUT method gives the highest energy accuracy and improves the relative energy resolution by 16\%, 22\%, and 21\%, respectively.

Notably, the results from the global linear calibration show slightly better energy accuracy than those from the pixel-wise linear calibration, but worse relative energy resolution.
This occurs because the global linear fit partially averages out the nonlinear deviations across pixels under a wide illumination, leading to a more consistent overall energy scale.
Conversely, the pixel-to-pixel gain variations degrade the relative energy resolution, as expected.

\begin{table*}
  \caption{Summary of the electron energy spectra in terms of the energy centroid ($E_{\rm centroid}$) and relative energy resolution (FWHM/$E_{\rm centroid}$) for different electron energies using different calibration methods.}
  \begin{tabular}{c|c|c|c|c|c|c}
    \bf{Fitted results} & \multicolumn{3}{c|}{\bf{$\boldsymbol{E_{\text{centroid}}}$ [keV]}} & \multicolumn{3}{c}{\bf{FWHM/$\boldsymbol{E_{\text{centroid}}}$ [\%]}} \\
    \hline
    \bf{Electron energy [keV]} & \bf{60} & \bf{80} & \bf{200} & \bf{60} & \bf{80} & \bf{200} \\ 
    \hline
    \bf{3D LUT calibration} & 59.59 & 79.75 & 200.23 & 4.07 & 3.39 & 3.03 \\
    \hline
    \bf{Pixel-wise linear gain} & 58.84 & 79.25 & 200.63 & 4.32 & 3.62 & 3.16 \\
    \hline
    \bf{Global linear gain} & 58.98 & 79.32 & 200.54 & 4.82 & 4.37 & 3.83 \\
  \end{tabular}
  \label{tab:electron_results}
\end{table*}

\subsection{Effects of 3D LUT calibration on the deep learning--based electron localization}

\label{sec:deepLearning}
\begin{figure*}
    \centering
    \includegraphics[width=0.45\linewidth]{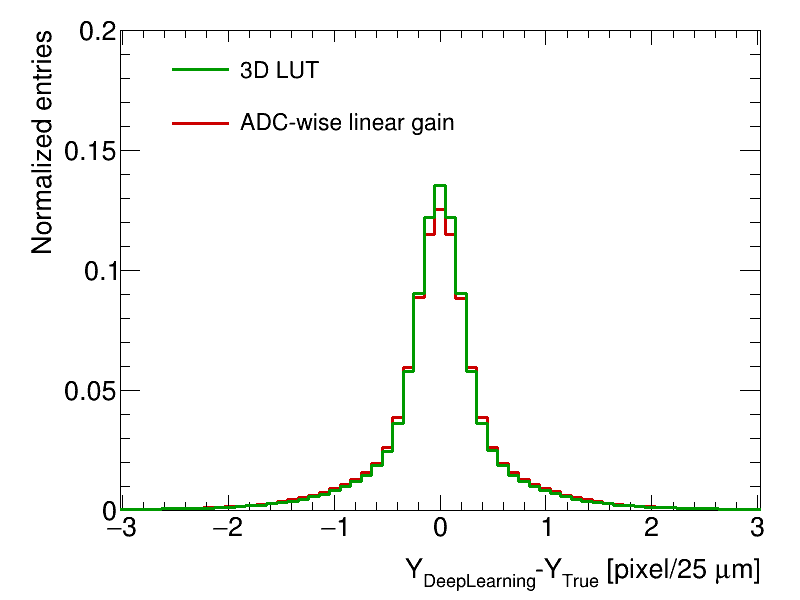}
    \caption{
      Residuals of the deep-learning predicted positions for 200~keV electrons along the Y axis, with the dataset processed using the ADC-wise (25$\times$200 pixels) linear calibration (red) and the 3D LUT calibration (green).
      Root mean square error (RMSE) improves from 0.609 pixel to 0.586 pixel, a 4\% improvement in spatial resolution.
    }
    \label{fig:deepLearning}
\end{figure*}

Deep learning--based electron localization has demonstrated subpixel accuracy using the MÖNCH detector \cite{EM_deepLearning2023}.
The performance of this method is sensitive to the training data quality, which makes it a suitable benchmark to quantify the impact of the proposed calibration method on data quality.

The raw data used to prepare training datasets were acquired on the same JEOL JEM-ARM200F NEOARM electron microscope at PSI, following the method described in \cite{EM_deepLearning2023}.
We reused the conventional linear calibration data from that work, which were performed ADC-wise (i.e., in blocks of 25$\times$200 pixels) to balance statistics with granularity.
The linear calibration was performed using X-ray fluorescence from multiple targets, including copper, selenium, zirconium, manganese, silver, and tin, for each 25$\times$200-pixel block.
For comparison, both the ADC-wise linear calibration and the proposed 3D LUT calibration were applied to the raw data.  
Electron clusters were formed using the same method as described in Section~\ref{sec:electron_results}.

Two neural network models with the same architecture as in \cite{EM_deepLearning2023} were trained separately using datasets obtained with different calibration methods.
70\% of the dataset was used for training, while the remaining 30\% was reserved for testing.
The residuals between the predicted and true electron positions for the test datasets of the two models are shown in Fig.~\ref{fig:deepLearning}.
The RMSE of the residuals improves from 0.609 pixel with the ADC-wise linear calibration to 0.586 pixel with the 3D LUT calibration.
The 3D LUT calibration yields a 4\% improvement in spatial resolution.  

\section{Discussion}
\label{sec:discussion}

\textbf{Linearity and uniformity of the pixel response}:
For both X-ray photons and electrons, the relative energy resolution improves from the global linear fit to the pixel-wise linear fit, and further to the 3D LUT calibration method.
This indicates that both the non-uniformity across different pixels and the nonlinearity of the pixel response contribute significantly to the degradation of the energy resolution.
To enhance the energy resolution performance of the MÖNCH detector, it is therefore necessary to adopt the 3D LUT calibration method to improve both the linearity of the pixel response and the uniformity across the pixel array.

\textbf{Bad pixel diagnosis}:
\begin{figure*}
    \centering
    \begin{subfigure}{0.3\linewidth}
      \centering
      \includegraphics[width=\linewidth]{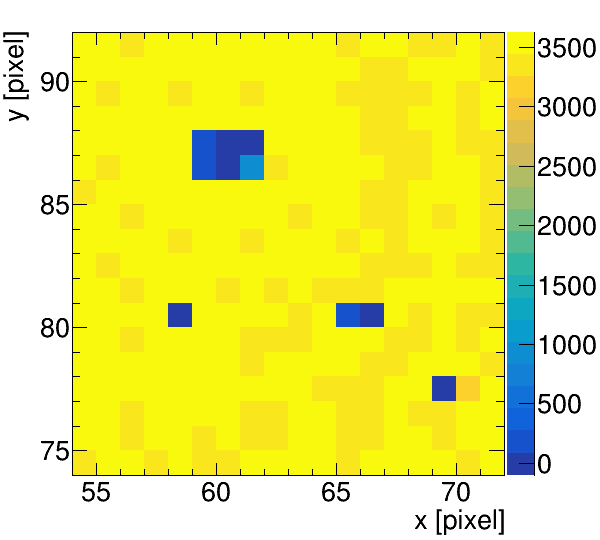}
      \subcaption{Bad electronic channels}
      \label{fig:diagnose_a}
    \end{subfigure} \ \ 
    \begin{subfigure}{0.3\linewidth}
      \centering
      \includegraphics[width=\linewidth]{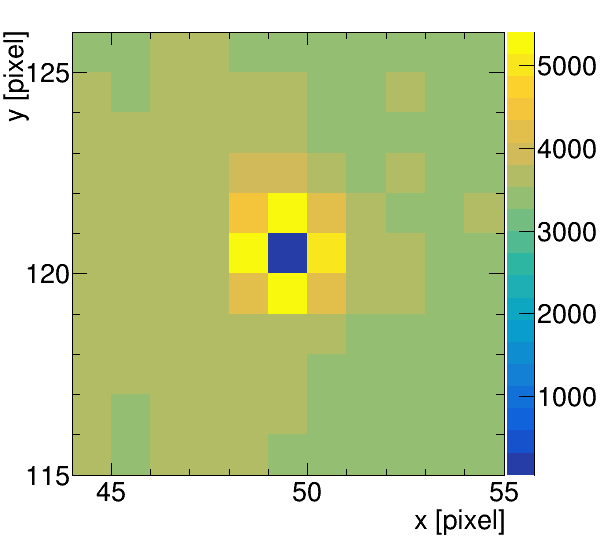}
      \subcaption{Unconnected bump-bonding}
      \label{fig:diagnose_b}
    \end{subfigure} \ \ 
    \begin{subfigure}{0.3\linewidth}
      \centering
      \includegraphics[width=\linewidth]{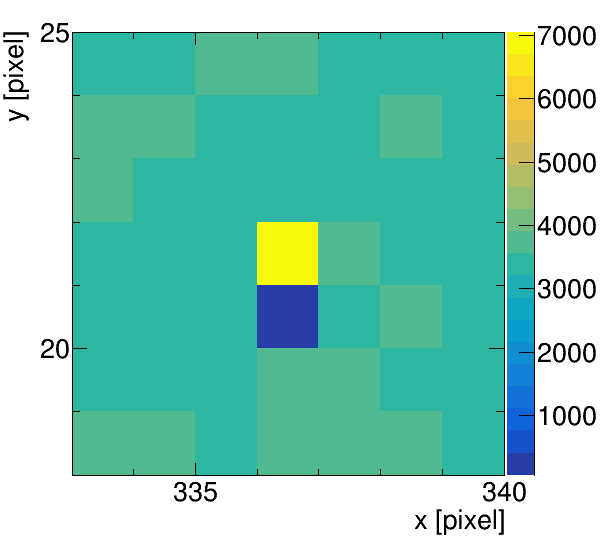}
      \subcaption{Short between indium bumps}
      \label{fig:diagnose_c}
    \end{subfigure} \ \ 

    \caption{
      Sub-regions of the pixel response to a given pulse amplitude, illustrating different types of bad pixel:
      (a) Bad electronic channel;
      (b) Unconnected bump-bonding, with the signal evenly shared among neighboring pixels;
      (c) Short between indium bumps, where one pixel outputs zero signal while an adjacent pixel collects approximately double the expected signal.
    }         
    \label{fig:diagnose}
\end{figure*}
Using this calibration approach, bad pixels can be efficiently identified and the pixel yield estimated.
Fig.~\ref{fig:diagnose} shows three representative types of different abnormal pixel behavior:
(a) pixels with faulty electronic channels, producing outputs close to zero across pulse amplitudes;
(b) a pixel with an unconnected bump-bonding, where the signal is symmetrically shared among neighboring pixels;
(c) short between indium bumps, where one pixel is inactive while an adjacent pixel collects approximately double the expected signal.
Since all pixels are pulsed simultaneously, this method is considerably faster than acquiring flat-field data over the entire detector to assess the pixel yield and diagnose bad pixels.
It is especially valuable for large detector modules, such as the planned full-size MÖNCH1.0 detector with 1024~$\times$~768 pixels for a single module \cite{Heymes2025Monch}.


\textbf{Practical usage of the 3D LUT calibration method}:
To be effectively integrated into the data analysis workflow, the 3D LUT calibration must be efficient in both time and memory usage.
The current calibration LUT is approximately 1.0~GB for the 400 $\times$ 400-pixel MÖNCH03 detector.
Given the low-frequency behavior of the nonlinearity, the size of the LUT can be further reduced by decreasing the number of bins.
As an example, a 22~MB LUT file with 36 bins (500 ADU/bin) is generated in the same way.
The 15~keV, 20~keV, and 25~keV photon results obtained via interpolation show consistent precision, with slightly larger biases (centroids at 14.96~keV, 20.00~keV, and 24.95~keV centroids).
In this way, the file size can remain manageable even for larger detectors with a few million pixels, while the energy resolution and accuracy are largely preserved.


Using scientific Python packages such as NumPy, converting ADU to energy is efficient if the LUT is buffered in memory.
For the 400 $\times$ 400-pixel MÖNCH03 detector, a speed test using buffered frames showed that the global linear calibration costs 0.012~ms per frame, while the 3D LUT calibration took 0.46~ms per frame without interpolation and 1.52~ms per frame with interpolation, respectively, using a single thread of an AMD Ryzen 9 7950X CPU.
The time consumptions of 3D LUT calibration was observed  not sensitive to the number of bins.
No significant increase in offline data analysis time is observed compared to conventional linear calibration methods, as reading data from disk and the cluster--finding algorithm typically dominate the total time.
This LUT-based calibration method also has the potential to be implemented on FPGA for real-time calibration.

\section{Conclusion}
\label{sec:conclusion}
We have developed an accurate and efficient calibration method for pixel-wise gain and nonlinearity calibration for charge-integrating hybrid pixel detectors, based on the backside pulsing technique.  
A three-dimensional lookup table (3D LUT) is generated for all pixels across the full dynamic range as the calibration file, mapping the pixel response to a calibrated linear energy scale.

The limitations of conventional linear calibration methods, including the pixel response nonlinearity and pixel--to--pixel non-uniformity, are addressed by this method.
Measurements with both photons and electrons demonstrate that this 3D LUT calibration method achieves significantly better energy accuracy and resolution than linear calibrations.
Deep learning--based electron localization also benefits from the improved calibration, with a 4\% improvement in spatial resolution.
The improved energy resolution and pixel--to--pixel uniformity are expected to further enhance subpixel interpolation applications in X-ray imaging and electron microscopy.
This method is also fast and practical, requiring only one hour to complete the calibration procedure in the current setup and exhibiting only limited dependence on the X-ray source.
The computational cost and memory usage are manageable even for large detectors.

In addition, we have discussed potential applications of this method for noise determination and pixel-yield diagnosis.
This calibration method is also applicable to other charge-integrating hybrid pixel detectors, such as the AGIPD and JUNGFRAU detectors; however, the level of improvement that can be obtained needs to be tested case by case.

\section*{Acknowledgments}
We acknowledge the usage of the instrumentation provided by the Electron Microscopy Facility at PSI and we thank the EMF team for their help and support. The authors thank Dr. Elisabeth Müller and Dr. Emiliya Poghosyan for their support with electron microscopy data acquisition.

We thank Benjamin Bejar Haro and Luis Barba Flores for their contributions to neural network development during the previous collaborative project.

The authors also gratefully acknowledge Marie Andrae and Arkadiusz Dawiec for their support during the beamtime,as well as the assistance provided by the METROLOGIE beamline staff and the detector group at the SOLEIL synchrotron.

K.A.P has received funding from the European Union’s Horizon 2020 research and innovation programme under the Marie Skłodowska-Curie grant agreement No 884104 (PSI-FELLOW-III-3i).

\bibliographystyle{elsarticle-num}

\bibliography{References}



\end{document}